\documentclass[journal]{IEEEtran}

\usepackage{placeins}
\usepackage{amsmath}
\usepackage{algorithm}
\usepackage{algorithmic}
\usepackage{booktabs}
\usepackage{multirow}
\usepackage[utf8]{inputenc} 
\usepackage[T1]{fontenc}    
\usepackage{hyperref}       
\usepackage{url}            
\usepackage{booktabs}       
\usepackage{amsfonts}       
\usepackage{nicefrac}       
\usepackage{microtype}      
\usepackage{xcolor}         
\usepackage{graphicx}
\usepackage{float}
\usepackage[numbers,sort&compress]{natbib}
\usepackage{caption} 

\newfloat{figtab}{htb}{fgtb}
\makeatletter
  \newcommand\figcaption{\def\@captype{figure}\caption}
  \newcommand\tabcaption{\def\@captype{table}\caption}
\makeatother

\def\BibTeX{{\rm B\kern-.05em{\sc i\kern-.025em b}\kern-.08em
    T\kern-.1667em\lower.7ex\hbox{E}\kern-.125emX}}

\title{Interpretable Fuzzy Modeling Reveals Population-Level Representation Differences in P300 Brain Computer Interfaces Across Neurodivergent and Neurotypical Cohorts}
\label{sec:title}
\author{
\IEEEauthorblockN{
Xiaowei Jiang\textsuperscript{1}\textsuperscript{*}, 
Sudong Shang\textsuperscript{2},
Adrian Wilkinson\textsuperscript{2},
Michael L. Platt\textsuperscript{3,4,5},
Da Xiao\textsuperscript{1},
Bening Cao\textsuperscript{1},
Thomas Do\textsuperscript{1}, 
\\
}

\IEEEauthorblockA{\textsuperscript{1}Computational Intelligence and Brain-Computer Interfaces lab, Australian AI Institute,\\ Faculty of Engineering and Information Technology, University of Technology Sydney, 2007, Australia\\
\textsuperscript{2}Department of Management, Griffith Business School, 4215, Australia\\
\textsuperscript{3}Department of Neuroscience, Perelman School of Medicine, University of Pennsylvania, Philadelphia, Pennsylvania, 19104, USA\\
\textsuperscript{4}Marketing Department, Wharton School of Business, University of Pennsylvania, Philadelphia, Pennsylvania, 19104, USA\\
\textsuperscript{5}Department of Psychology, School of Arts and Sciences, University of Pennsylvania, Philadelphia, Pennsylvania, 19104, USA\\
}

}

\begin{document}
\maketitle

\begin{abstract}
P300-based brain-computer interfaces (BCIs) are widely used for communication, but population heterogeneity may alter the neural patterns available for decoding. Prior work has mainly examined such differences at the signal or performance level, while the representation structure learned by the decoder remains underexplored. In this study, we propose an interpretable fuzzy spatiotemporal framework for P300 classification and use it to analyze population-level differences across amyotrophic lateral sclerosis (ALS), autism (AUT), and neurotypical (NT) cohorts. The model employs spatial and temporal fuzzy filters with learnable prototypes, enabling both classification and reconstruction of cohort-specific fuzzy centers. Experiments were conducted on ALS and NT subsets from bigP3BCI and on the BCIAUT-P300 benchmark in a within-subject setting. The proposed model achieved competitive performance against multiple deep learning baselines. More importantly, the reconstructed fuzzy centers revealed systematic cohort-dependent differences in waveform morphology and representation geometry. Point-wise statistical analysis identified significant temporal differences between cohorts, including intervals overlapping with the canonical P300 window, and low-dimensional embeddings showed partially separated cohort-specific prototype organizations. These results suggest that population heterogeneity in P300-BCI is reflected not only in decoding performance but also in the discriminative structure learned by the model. The proposed framework provides an interpretable route toward population-aware P300-BCI analysis and design.
\end{abstract}


\section{Introduction}

Brain computer interfaces based on the P300 event-related potential remain one of the most established noninvasive approaches for communication and control~\cite{Allison2020}, particularly for users with limited motor output~\cite{Sellers2006, Nijboer2008, Donchin2009} and low training tolerance~\cite{Elshout2009}. Neurophysiologically, the P300, especially the P3b component, reflects task-relevant stimulus evaluation, attentional allocation, and context updating, making it well suited for oddball-based assistive systems~\cite{Polich2007,McFarland2011}. Consequently, P300-BCI has become a major paradigm in translational BCI research.

A major obstacle to practical deployment, however, is that discriminative neural responses are often not sufficiently stable across users or recording sessions. Prior EEG and BCI studies have shown that both inter-subject and intra-subject variability can induce distribution shift, thereby weakening model generalization and transferability across subjects and sessions~\cite{Saha2019,Huang2023}. This instability also increases calibration demands in practical BCI deployment, as substantial performance degradation can occur across sessions or datasets when models are transferred without adaptation~\cite{Xu2020}. Although transfer learning may partially alleviate this problem and markedly reduce calibration time~\cite{Wei2018}, variability remains a fundamental challenge for robust real-world EEG decoding. This issue is likely to be even more pronounced in clinical and neurodiverse populations, where differences in sensory processing, attention, cognition, and disease burden may further alter the morphology and separability of event-related responses.

This issue is particularly relevant for autistic and neurotypical populations. Over the past two decades, ERP studies have consistently suggested atypical sensory and attentional processing in autism. A meta-analysis of 32 studies reported significantly reduced P3b amplitude in ASD, especially in oddball paradigms and over parietal sites~\cite{Cui2017}. More recent work likewise supports attenuated or delayed P300-related responses under attentionally demanding conditions~\cite{Lassen2025}. In parallel, P300-BCI research in autism has advanced through joint-attention and social cueing paradigms, with datasets such as BCIAUT-P300 enabling systematic cross-subject evaluation~\cite{Simoes2020}. These studies suggest that P300 detection in autistic participants may be more difficult because the signal is often weaker, noisier, or temporally shifted relative to neurotypical populations~\cite{Peketi2023}.

A related translational challenge arises in amyotrophic lateral sclerosis (ALS). ALS is a key target population for P300-BCI because progressive motor decline can severely impair communication while ERP-based visual selection may remain available~\cite{McCane2014,Cipresso2012}. Yet ALS is also heterogeneous. Even when overall P300-BCI accuracy appears comparable to controls, the underlying ERP morphology may differ in amplitude, latency, and scalp distribution~\cite{McCane2015,Riccio2018}. In addition, cognitive and behavioral impairment can degrade control signals and reduce BCI performance~\cite{Geronimo2016,Riccio2018,Vuckovic2014}. Recent evidence further suggests that such population differences may materially affect transfer learning in P300-BCI. In an ALS-focused study based on bigP3BCI, generic EEGNet models pretrained on non-ALS data performed poorly, whereas models pretrained on ALS data performed comparably to or better than user-specific baselines~\cite{Lin2025PopulationTransfer}. The authors also noted that this degradation may reflect not only population differences, but also associated shifts in experimental conditions such as grid size and inter-stimulus interval~\cite{Lin2025PopulationTransfer}. Together, these findings suggest that in both autism and ALS, population differences are unlikely to be captured fully by average decoding accuracy alone. However, existing work has largely remained at the level of transfer performance, rather than examining what discriminative structure the decoder actually learns across populations.

Despite this background, most prior studies examine population differences at either the signal level, such as ERP amplitude, latency, or topography, or the performance level, such as accuracy and usability. What remains largely missing is a representation-level account of what the decoder actually learns as group-specific discriminative structure. This gap is important in heterogeneous populations, where reduced performance may reflect not only lower signal-to-noise ratio but also systematic changes in the geometry of target and non-target representations.

This limitation also highlights the importance of interpretability. In healthcare-facing AI, predictive accuracy alone is insufficient if the model cannot provide transparent and clinically meaningful reasoning~\cite{Sadeghi2024, Abbas2025}. This is particularly relevant for EEG-based BCI, where black-box decoders may obscure whether decisions are driven by robust neurophysiological structure or dataset-specific shortcuts. Fuzzy systems are well suited to this problem because they naturally model uncertainty, overlap, and graded class membership, which are intrinsic to noisy and heterogeneous neural data~\cite{jiang2024ifuzzytl, 10849810, cao2025emdfuzzy, jiang2025interpretable}, especially medical data~\cite{Nauck1999}.

In this study, we address these issues using an interpretable fuzzy modeling framework applied separately to AUT, ALS, and NT P300 data. Beyond decoding accuracy, we analyze the learned fuzzy centers as population-level global knowledge to characterize how discriminative P300 representations differ across groups. This allows us to ask whether reduced decoding performance in non-NT or clinical cohorts reflects only poorer classification, or also a systematic reorganization of the learned centers that separate target from non-target responses.

Our contributions are threefold. First, we provide a unified comparison of P300 decoding across AUT, ALS, and NT cohorts within the same fuzzy framework. Second, we move beyond conventional performance reporting by analyzing learned centers as interpretable population-level representations. Third, we show that reduced P300 recognition performance in clinically and neurocognitively distinct populations is accompanied by systematic differences in learned center structure, suggesting that population heterogeneity is expressed at the representation level rather than merely as additional noise. Together, these findings support population-aware and interpretable P300-BCI design and position fuzzy center analysis as a bridge between ERP neuroscience and translational neuroengineering.

\section{Methods and Materials}

\subsection{P300 Paradigm and Learning Objective}
\label{sec:p300_principle}

The present study focuses on P300-based brain-computer interface decoding under a visual oddball setting. In this paradigm, infrequent task-relevant stimuli elicit a positive event-related potential that typically emerges several hundred milliseconds after stimulus onset and is most prominent over parietal regions. The P300, particularly the P3b component, is widely linked to stimulus evaluation, attentional allocation, and context updating, which makes it a standard signal basis for assistive BCI systems~\cite{Mattout2015CoadaptationP300}.

Unlike steady-state paradigms that are defined by stimulus frequency~\cite{Norcia2015}, the P300 task is fundamentally an event-locked discrimination problem. Therefore, the goal here is not to identify a stimulation frequency, but to determine whether a recorded EEG epoch corresponds to a target or a non-target event. Let
\[
\mathbf{X} \in \mathbb{R}^{C \times T}
\]
denote a single EEG epoch, where \(C\) is the number of electrodes and \(T\) is the number of sampled time points. The associated label is denoted by
\[
z \in \{0,1\},
\]
where \(z=1\) indicates a target trial and \(z=0\) indicates a non-target trial.

This formulation is well suited to fuzzy modeling because P300 responses are not strictly identical across individuals or cohorts. Instead, they exhibit graded variation in amplitude, latency, and spatial distribution, making hard partitioning less appropriate than soft rule-based assignment~\cite{Picton1992}.

\subsection{Problem Formulation}

Assume that the training set is collected from \(N\) subjects:
\[
\mathcal{D} = \{\mathcal{D}_1,\mathcal{D}_2,\ldots,\mathcal{D}_N\},
\]
where the data of subject \(n\) are written as
\[
\mathcal{D}_n = \{(\mathbf{X}^{(i)}_n, z^{(i)}_n)\}_{i=1}^{M_n}.
\]
Each trial \(\mathbf{X}^{(i)}_n \in \mathbb{R}^{C \times T}\) is an EEG epoch aligned to stimulus onset, and \(z^{(i)}_n\) is its binary label.

We study cross-subject P300 decoding in a leave-one-subject-out setting. For each fold, data from one subject are held out for testing, while the remaining subjects are used for model training. This setting evaluates whether subject-invariant yet interpretable decision structure can be learned from heterogeneous populations.

In addition to prediction, we aim to characterize the global discriminative representations learned from each cohort. For this purpose, the proposed model is designed not only to output class probabilities, but also to learn fuzzy prototypes that can be traced back to the input domain for interpretation.

\subsection{Dataset Description}

We evaluated the proposed framework on P300 datasets from neurodivergent and neurotypical cohorts. The neurodivergent group consisted of an ALS cohort and an autism cohort. The ALS data were selected from the bigP3BCI collection, specifically studies L and N. Study L includes 11 participants with ALS recorded in 1 session, and study N includes 8 participants with ALS recorded in 2 sessions. These studies employ checkerboard-based P300 paradigms, with study L additionally including row-column and color-checkerboard conditions \cite{Mainsah2024bigP3BCI,Ryan2018ALS,Clements2016Dry}.

The autism cohort was taken from the BCIAUT-P300 benchmark dataset, which contains 15 autistic participants recorded across 7 sessions, yielding 105 sessions in total. BCIAUT-P300 was released as a multi-session, multi-subject benchmark resource for P300-based BCI research in autism \cite{Simoes2020}.

The neurotypical cohort was selected from bigP3BCI studies E, Q, S1, and S2. According to the bigP3BCI dataset documentation, these studies comprise 8, 36, 10, and 24 neurotypical participants, respectively, and mainly use checkerboard-based paradigms \cite{Mainsah2024bigP3BCI}.

Overall, this dataset composition enabled us to compare learned P300 representations across clinically and neurocognitively distinct populations, including ALS, autism, and neurotypical participants, within a unified analysis framework \cite{Mainsah2024bigP3BCI,Simoes2020}.

\begin{figure*}
    \centering
    \includegraphics[width=1\linewidth]{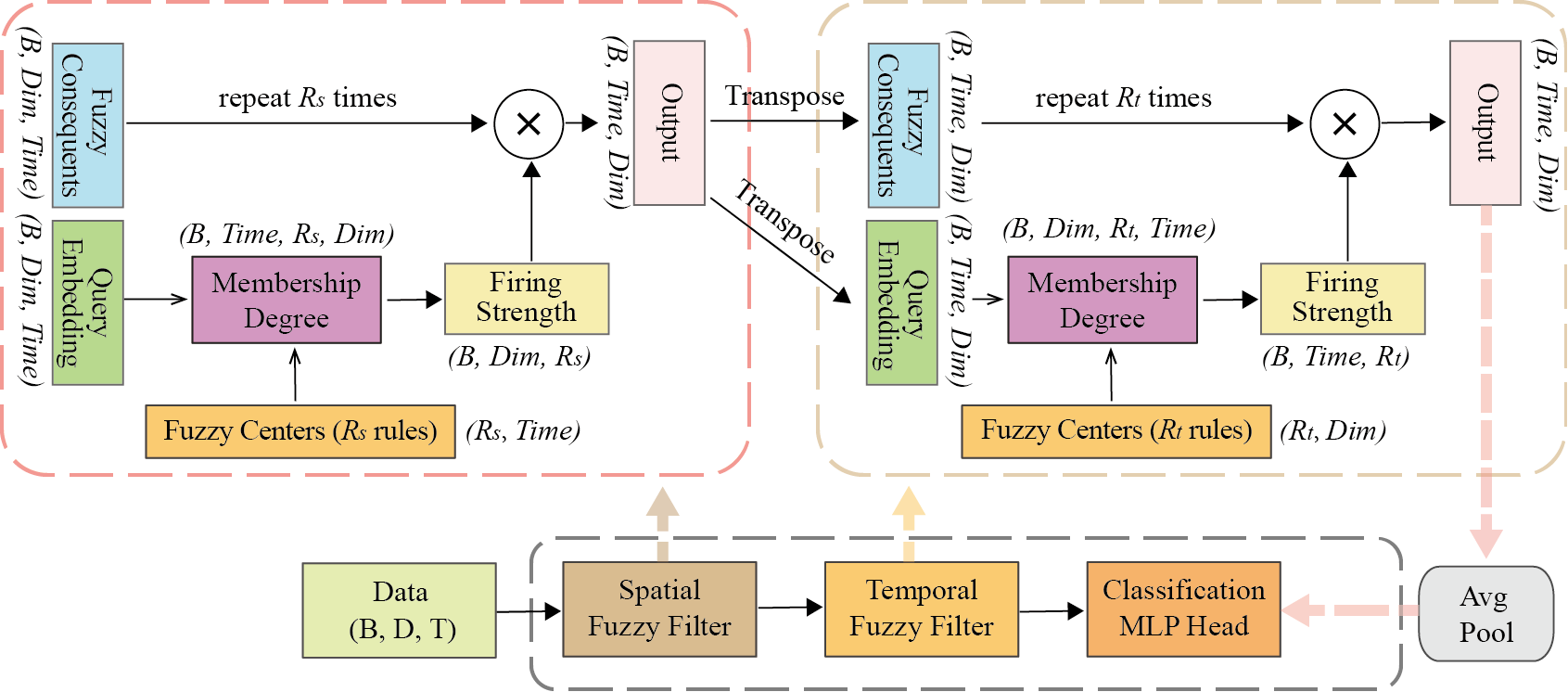}
    \caption{Overall architecture of the proposed interpretable fuzzy P300 decoder. The model consists of a spatial fuzzy filter, a temporal fuzzy filter, and a classification MLP head. Given an input EEG epoch of shape \((B, D, T)\), where \(B\) denotes batch size, \(D\) denotes the number of channels, and \(T\) denotes the number of time points, the spatial fuzzy filter first performs rule-guided aggregation across the channel dimension using \(R_s\) learnable fuzzy centers. The resulting representation is then processed by the temporal fuzzy filter, which applies an analogous fuzzy inference mechanism across the temporal dimension using \(R_t\) learnable fuzzy centers.}
    \label{fig:model_structure}
\end{figure*}

\subsection{Interpretable Fuzzy Spatiotemporal Encoder}

The proposed model, termed iFuzzyP300 here for exposition, consists of a spatial fuzzy module, a temporal fuzzy module, and a lightweight classifier. The overall design preserves the original architecture of iFuzzyTL but is reformulated for event-related P300 decoding.

The key idea is to map each input trial into a latent space, evaluate its similarity to a set of learnable fuzzy prototypes, and then use the resulting rule activations to guide feature aggregation. This design is particularly suitable for P300 analysis because the discriminative structure is shaped by soft variations in waveform morphology rather than by rigid class boundaries.

\subsubsection{Fuzzy Rule Activation in Latent Space}

Given an input representation \(\mathbf{h} \in \mathbb{R}^{Q}\), the activation of rule \(k\) is defined through a Gaussian prototype function:
\begin{equation}
\rho_k(\mathbf{h}) =
\exp\left(
-\sum_{q=1}^{Q}
\frac{(h_q-\mu_{kq})^2}{2\delta_{kq}^2}
\right),
\end{equation}
where \(\mu_{kq}\) and \(\delta_{kq}\) are the learnable center and width of rule \(k\) along latent dimension \(q\), respectively.

The normalized contribution of each rule is then computed as
\begin{equation}
\pi_k(\mathbf{h}) =
\frac{\rho_k(\mathbf{h})}
{\sum_{j=1}^{K}\rho_j(\mathbf{h})},
\label{eq:new_rule_norm}
\end{equation}
where \(K\) denotes the total number of fuzzy rules.

Equation~(\ref{eq:new_rule_norm}) yields a soft partition of the latent representation space. Instead of forcing each trial into a single rigid cluster, the model expresses each trial as a graded combination of rule responses. This is desirable for P300 decoding because event-related responses often overlap substantially between target and non-target classes, especially across heterogeneous subjects and clinical cohorts.

\subsubsection{Rule-Guided Spatial Aggregation}

Let the input epoch be denoted by
\[
\mathbf{X} \in \mathbb{R}^{C \times T}.
\]
For spatial modeling, the signal is processed across the channel dimension so that the model can learn cohort-level channel contribution patterns. A rule-specific query projection and value projection are first applied:
\begin{equation}
\mathbf{q}^{(s)}_k = \mathbf{W}^{(s,Q)}_k \mathbf{X}, 
\qquad
\mathbf{v}^{(s)}_k = \mathbf{W}^{(s,V)}_k \mathbf{X},
\end{equation}
where \(\mathbf{W}^{(s,Q)}_k\) and \(\mathbf{W}^{(s,V)}_k\) are learnable spatial projection parameters for rule \(k\).

The projected query is then compared with the corresponding fuzzy prototype to obtain a normalized rule weight \(\pi^{(s)}_k\). The output of the spatial fuzzy module is given by
\begin{equation}
\mathbf{H}^{(s)} =
\sum_{k=1}^{K_s}
\pi^{(s)}_k \odot \mathbf{v}^{(s)}_k,
\label{eq:new_spatial}
\end{equation}
where \(K_s\) is the number of spatial rules and \(\odot\) denotes elementwise weighting with broadcasting as needed.

This module can be interpreted as a prototype-guided spatial encoder. Rather than assigning fixed channel weights, it adaptively emphasizes spatial patterns that are most compatible with the learned fuzzy prototypes.

\subsubsection{Rule-Guided Temporal Aggregation}

The output of the spatial module is then transposed and passed to the temporal module to model event-related waveform structure over time. For each temporal rule \(m\),
\begin{equation}
\mathbf{q}^{(t)}_m = \mathbf{W}^{(t,Q)}_m \mathbf{H}^{(s)},
\qquad
\mathbf{v}^{(t)}_m = \mathbf{W}^{(t,V)}_m \mathbf{H}^{(s)},
\end{equation}
where \(\mathbf{W}^{(t,Q)}_m\) and \(\mathbf{W}^{(t,V)}_m\) are learnable temporal projection parameters.

Analogously, temporal rule activations are computed via Gaussian prototype matching and normalized across rules. The temporal output is then obtained as
\begin{equation}
\mathbf{H}^{(t)} =
\sum_{m=1}^{K_t}
\pi^{(t)}_m \odot \mathbf{v}^{(t)}_m,
\label{eq:new_temporal}
\end{equation}
where \(K_t\) is the number of temporal rules.

This second stage is important for P300 analysis because class discrimination depends not only on where information appears over the scalp, but also on when the relevant event-related deflection emerges within the epoch.

\subsubsection{Prototype Reconstruction for Interpretability}

To improve interpretability, the learned fuzzy prototypes in latent space are projected back to the signal domain. Suppose a latent prototype is denoted by \(\mathbf{c}\), and the corresponding linear projection is
\begin{equation}
\mathbf{u} = \mathbf{P}\mathbf{x} + \mathbf{b},
\end{equation}
where \(\mathbf{x}\) is an input-domain representation, \(\mathbf{P}\) is the projection matrix, and \(\mathbf{b}\) is the bias term.

When \(\mathbf{P}\) is invertible, the corresponding signal-domain pattern can be recovered directly as
\begin{equation}
\mathbf{x} = \mathbf{P}^{-1}(\mathbf{u}-\mathbf{b}).
\end{equation}
For non-square or rank-deficient cases, we use the Moore-Penrose pseudoinverse:
\begin{equation}
\mathbf{x} = \mathbf{P}^{+}(\mathbf{u}-\mathbf{b}).
\end{equation}

This reconstruction step does not introduce an auxiliary decoder. Instead, it maps latent prototypes back to the original signal space using the learned linear projection itself, allowing the model to expose what type of waveform pattern each fuzzy rule represents. This is aligned with the general goal of interpretable fuzzy systems, which is to retain readable rule structure rather than only predictive performance~\cite{jiang2024ifuzzytl, 10849810}.

\subsubsection{Classification Head and Optimization}

The output of the temporal module is flattened and fed into a two-layer multilayer perceptron with ReLU activation and dropout. Let the predicted probability of the target class for trial \(i\) be denoted by \(\hat{p}_i\). The model is trained using binary cross-entropy:
\begin{equation}
\mathcal{L}_{\mathrm{cls}}
=
-\frac{1}{M}
\sum_{i=1}^{M}
\left[
z_i \log \hat{p}_i + (1-z_i)\log(1-\hat{p}_i)
\right],
\end{equation}
where \(M\) is the number of training trials in the current batch.

The overall training objective is to learn a subject-general decision function while preserving interpretable rule structure through the fuzzy prototype parameterization.

The proposed architecture is particularly suitable for P300 decoding for two reasons. First, P300 discrimination depends on coupled spatial and temporal structure rather than on a single spectral signature, making a dual spatiotemporal encoder more appropriate than frequency-oriented modeling. Second, because P300 morphology varies across individuals in amplitude, latency, and scalp distribution, fuzzy prototype matching provides a natural mechanism for handling graded inter-subject variation instead of enforcing hard feature partitions~\cite{Bhattacharjee2017,SalimiKhorshidi2008}.

\subsection{Within-Subject Evaluation Protocol}

We evaluated the model using the predefined within-subject training and test split provided by each dataset. For each subject, the original training set was further divided into training and validation subsets with a ratio of 8:2, while the original test set was kept untouched and used only for final performance evaluation.

This evaluation strategy assesses the model under a subject-specific decoding setting, where training and testing are performed within the same individual while preserving a strict separation between model development and final testing. Such a protocol is appropriate for P300-BCI analysis, as it reflects the common practical scenario in which a personalized decoder is trained using calibration data from the target user and then evaluated on unseen trials from the same subject.

\subsection{Training Strategy}

For each subject, the predefined training partition of the dataset was further divided into training and validation subsets using an 8:2 ratio. The independent test partition provided by the dataset was kept untouched and used only for final evaluation.

All models were trained using the AdamW optimizer with an initial learning rate scaled from a base value of \(1.5 \times 10^{-3}\) according to batch size. The batch size was set to 256 for training, validation, and testing. Optimization was performed for at most 800 epochs using a warmup cosine decay learning-rate schedule, with 80 warmup epochs and a weight decay of 0.05.

Model selection was based on the validation performance. Early stopping was applied with a patience of 50 epochs and a minimum improvement threshold of 0.0005. The checkpoint with the best validation performance was retained for final testing.

To reduce irrelevant physiological and environmental noise, the EEG signals were band-pass filtered between 0.1 and 30 Hz and notch filtered at 50 Hz. Each epoch was extracted from \(-0.195\) s to \(0.8\) s relative to stimulus onset.

\subsection{Performance Metrics}

To evaluate decoding performance, we used four metrics: accuracy, F1 score, precision, and recall. Accuracy measures the overall proportion of correctly classified trials. Precision quantifies the proportion of trials predicted as target that were truly target trials, whereas recall measures the proportion of true target trials that were correctly detected. The F1 score, defined as the harmonic mean of precision and recall, provides a balanced summary of detection performance under class imbalance.

These metrics were selected because P300 classification is typically formulated as an imbalanced binary problem, where non-target trials substantially outnumber target trials. In such settings, relying on accuracy alone may provide an incomplete view of classifier behavior, making precision-, recall-, and F1-based evaluation necessary.

\section{Results}

\subsection{Comparison with Baseline Models Across Cohorts}

Table~\ref{tab:fuzzy_vs_baselines} summarizes the performance of the proposed fuzzy model and baseline deep learning models across ALS, AUT, and NT cohorts. Overall, the fuzzy model achieved competitive performance across all cohorts and studies, while showing particularly stable behavior in the neurodivergent cohorts. In the AUT cohort, the proposed model remained more robust than several baseline architectures, especially compared with EEGNet and EEGformer, which exhibited marked performance degradation. In the ALS and NT cohorts, the fuzzy model was generally comparable to strong convolutional baselines, although some baseline models achieved higher scores on specific studies. These findings indicate that the proposed fuzzy framework provides a competitive and stable alternative for P300 decoding across heterogeneous populations.

Importantly, performance alone does not reveal whether different cohorts share the same discriminative structure. We therefore next examined the learned fuzzy centers to characterize population-level representation differences beyond aggregate decoding accuracy.

\begin{table*}[t]
\centering
\caption{Performance comparison between the proposed fuzzy model and baseline deep learning models across cohorts and studies. Results are reported as mean (standard deviation). Significance markers for baseline models indicate paired comparisons against the fuzzy model: * $p<0.05$, ** $p<0.01$, *** $p<0.001$.}
\label{tab:fuzzy_vs_baselines}
\resizebox{\textwidth}{!}{
\begin{tabular}{lllllll}
\toprule
Cohort & Study & Model & Accuracy & F1 & Precision & Recall \\
\midrule

\multirow{5}{*}{ALS}
& \multirow{5}{*}{Study L}
& Fuzzy                  & 86.81 (2.69) & 82.53 (3.71) & 80.59 (4.92) & 86.81 (2.69) \\
& & Conformer-BIGP3BCI      & 86.02 (3.77) & 81.51 (4.50) & 77.88 (5.66)*   & 86.02 (3.77) \\
& & DeepConvNet-BIGP3BCI    & 86.69 (4.21) & 86.40 (4.55)** & 86.49 (5.07)** & 86.69 (4.21) \\
& & EEGNet-BIGP3BCI         & 87.87 (3.19)* & 85.67 (4.36)* & 85.29 (4.70)* & 87.87 (3.19)* \\
& & ShallowConvNet-BIGP3BCI & 88.60 (3.40)* & 86.97 (4.02)** & 86.40 (4.39)** & 88.60 (3.40)* \\
\midrule

\multirow{5}{*}{ALS}
& \multirow{5}{*}{Study N}
& Fuzzy                  & 88.46 (1.44) & 84.59 (2.25) & 83.00 (3.79) & 88.46 (1.44) \\
& & Conformer-BIGP3BCI      & 87.31 (5.52) & 84.07 (3.82) & 81.58 (4.40) & 87.31 (5.52) \\
& & DeepConvNet-BIGP3BCI    & 87.47 (2.18) & 87.28 (2.75)* & 87.40 (3.61)* & 87.47 (2.18) \\
& & EEGNet-BIGP3BCI         & 89.41 (3.84) & 87.73 (4.35) & 87.18 (4.65)* & 89.41 (3.84) \\
& & ShallowConvNet-BIGP3BCI & 89.63 (1.67) & 87.54 (2.94)* & 86.87 (3.44)* & 89.63 (1.67) \\
\midrule

\multirow{6}{*}{AUT}
& \multirow{6}{*}{BCIAUT}
& Fuzzy              & 86.24 (1.62) & 81.84 (0.90) & 79.80 (2.00) & 86.24 (1.62) \\
& & Conformer           & 85.56 (4.46) & 80.92 (1.91)*** & 77.08 (1.11)*** & 85.56 (4.46) \\
& & DeepConvNet         & 86.66 (2.14) & 86.12 (2.74)*** & 85.99 (3.37)*** & 86.66 (2.14) \\
& & EEGNet              & 74.14 (20.13)*** & 75.05 (15.09)*** & 82.92 (4.08)*** & 74.14 (20.13)*** \\
& & EEGformer           & 33.75 (0.00) & 40.06 (0.00) & 78.33 (0.00) & 33.75 (0.00) \\
& & ShallowConvNet      & 87.20 (1.82)*** & 85.54 (2.22)*** & 84.85 (2.59)*** & 87.20 (1.82)*** \\
\midrule

\multirow{5}{*}{NT}
& \multirow{5}{*}{Study E}
& Fuzzy                  & 92.45 (2.33) & 89.23 (3.62) & 87.04 (4.76) & 92.45 (2.33) \\
& & Conformer-BIGP3BCI      & 91.89 (0.33) & 88.63 (2.11) & 85.77 (4.11) & 91.89 (0.33) \\
& & DeepConvNet-BIGP3BCI    & 91.91 (2.81) & 92.07 (2.95) & 92.38 (3.25) & 91.91 (2.81) \\
& & EEGNet-BIGP3BCI         & 92.99 (3.74) & 91.73 (4.52) & 91.49 (4.94) & 92.99 (3.74) \\
& & ShallowConvNet-BIGP3BCI & 93.79 (2.84) & 92.77 (3.60) & 92.50 (3.99) & 93.79 (2.84) \\
\midrule

\multirow{5}{*}{NT}
& \multirow{5}{*}{Study Q}
& Fuzzy                  & 91.49 (0.68) & 88.35 (1.08) & 87.39 (2.57) & 91.49 (0.68) \\
& & Conformer-BIGP3BCI      & 91.37 (1.02) & 87.67 (0.34) & 84.55 (1.08)* & 91.37 (1.02) \\
& & DeepConvNet-BIGP3BCI    & 90.09 (1.20)** & 90.28 (1.30)* & 90.80 (1.97)* & 90.09 (1.20)** \\
& & EEGNet-BIGP3BCI         & 89.42 (11.46) & 88.73 (8.45) & 90.27 (2.89) & 89.42 (11.46) \\
& & ShallowConvNet-BIGP3BCI & 92.09 (0.90) & 90.75 (1.54)* & 90.27 (1.93)* & 92.09 (0.90) \\
\midrule

\multirow{5}{*}{NT}
& \multirow{5}{*}{Study S1}
& Fuzzy                  & 92.06 (1.86) & 90.96 (2.36) & 90.47 (2.70) & 92.06 (1.86) \\
& & Conformer-BIGP3BCI      & 90.92 (1.80) & 87.51 (0.46) & 85.09 (2.28) & 90.92 (1.80) \\
& & DeepConvNet-BIGP3BCI    & 93.45 (2.74) & 94.07 (2.10) & 95.35 (0.93) & 93.45 (2.74) \\
& & EEGNet-BIGP3BCI         & 92.78 (7.57) & 92.90 (5.57) & 93.61 (2.34) & 92.78 (7.57) \\
& & ShallowConvNet-BIGP3BCI & 95.54 (1.00) & 95.25 (1.20) & 95.25 (1.23) & 95.54 (1.00) \\
\midrule

\multirow{5}{*}{NT}
& \multirow{5}{*}{Study S2}
& Fuzzy                  & 90.22 (5.51) & 88.64 (3.98) & 88.18 (2.82) & 90.22 (5.51) \\
& & Conformer-BIGP3BCI      & 91.42 (0.70) & 87.75 (0.47) & 84.52 (1.31) & 91.42 (0.70) \\
& & DeepConvNet-BIGP3BCI    & 91.75 (2.76) & 92.50 (2.32)** & 93.87 (1.92) & 91.75 (2.76) \\
& & EEGNet-BIGP3BCI         & 85.81 (14.91) & 87.64 (10.95) & 91.51 (4.51)* & 85.81 (14.91) \\
& & ShallowConvNet-BIGP3BCI & 94.51 (1.75)* & 93.89 (2.18)* & 93.83 (2.34) & 94.51 (1.75)* \\
\bottomrule
\end{tabular}
}
\end{table*}

\subsection{Interpretable Center Analysis Across Cohorts}

To investigate whether cohort differences were reflected at the representation level rather than only in aggregate decoding performance, we further examined the learned fuzzy centers from the ALS, AUT, and NT models. Figure~\ref{fig:center_interpretability} shows the reconstructed center waveforms for five representative rules together with point-wise statistical comparisons and low-dimensional embeddings of the corresponding centers.

Several consistent patterns can be observed. First, the reconstructed centers were not identical across cohorts. In multiple rules, the AUT centers showed visibly larger deviations in waveform morphology relative to the ALS and NT centers, especially in the post-stimulus interval. By contrast, the ALS and NT centers were often more similar to each other in overall shape, although rule-specific differences were still present.

Second, these cohort effects were not limited to qualitative visual differences. The point-wise statistical results above each waveform panel indicate that significant between-group differences occurred across multiple temporal segments, including intervals overlapping with the expected P300-related response window. In particular, repeated significant differences were observed for AUT versus NT and ALS versus NT, suggesting that the learned prototypes capture systematic cohort-dependent variations in event-related structure rather than random fluctuations.

Third, the low-dimensional embeddings of the reconstructed centers further support this interpretation. For each rule, the centers from ALS, AUT, and NT formed partially separated clusters in the embedding space, indicating that the fuzzy model learned cohort-specific prototype organizations. This separation was especially clear for AUT, whose centers tended to occupy a distinct region from NT, while ALS often showed an intermediate or partially overlapping configuration depending on the rule.

Taken together, these results suggest that population heterogeneity in P300-BCI is expressed not only in performance differences, but also in the geometry of the learned discriminative prototypes. The proposed fuzzy framework therefore provides an interpretable means of exposing how target and non-target decision structure is reorganized across neurotypical, autistic, and ALS cohorts.

\begin{figure*}[t]
    \centering
    \includegraphics[width=0.8\linewidth]{Figure/interpretability.png}
    \caption{Interpretable analysis of reconstructed fuzzy centers across cohorts. Left: reconstructed center waveforms for five representative rules learned from the ALS, AUT, and NT cohorts. Horizontal colored bars above each panel indicate time intervals showing significant effects from point-wise one-way ANOVA and post-hoc pairwise comparisons between ALS and NT, and between AUT and NT. The black vertical line marks stimulus onset, and the red window highlights the approximate P300-related interval. Right: low-dimensional embeddings of the reconstructed centers for each rule, showing cohort-specific organization in representation space. Across rules, the learned centers exhibit systematic waveform and geometric differences between ALS, AUT, and NT, supporting the presence of population-level representational heterogeneity.}
\label{fig:center_interpretability}
\end{figure*}

\section{Discussion}

This study examined P300 decoding across ALS, AUT, and NT cohorts using an interpretable fuzzy framework. The main finding is that population heterogeneity was reflected not only in decoding performance, but also in the structure of the learned prototypes. Although the proposed fuzzy model was generally competitive with strong deep learning baselines across cohorts, its main advantage in the present setting was interpretability: the reconstructed centers revealed systematic cohort-dependent differences in waveform morphology and representation geometry that would not be visible from accuracy alone.

These findings are consistent with prior evidence that P300 responses vary across neurotypical, autistic, and ALS populations. Earlier work has reported atypical P300 amplitude and latency in autism, as well as broader ERP abnormalities related to sensory and higher-order cognitive processing in ASD \cite{Cui2017,Jeste2009}. In ALS, prior P300-BCI work has shown that even when overall communication accuracy may remain comparable to controls, ERP morphology can still differ in component amplitude, latency, and scalp distribution \cite{McCane2015}. Our results extend this literature by showing that such differences can also be expressed at the representation level learned by the decoder itself.

A second implication is methodological. Prior work on EEG-based and P300-based transfer learning has emphasized that subject- and domain-related variability remains a major obstacle for generalized BCI decoding, motivating alignment and adaptation methods across subjects, sessions, and datasets \cite{Wu2022,Altindis2023}. Relatedly, the BCIAUT-P300 benchmark also highlights the practical challenges posed by inter-subject variability in an autism cohort \cite{Simoes2020}. The present results provide a possible interpretability-oriented explanation for that phenomenon: if different cohorts occupy different prototype organizations in latent and reconstructed signal space, then models trained on one population may not transfer well to another even when the task is nominally the same. In this sense, the learned fuzzy centers may serve as an interpretable bridge between performance variation and population-specific signal organization.

While these findings originate from EEG-based decoding contexts, they may have broader implications for workplace research and practice. In particular, the possibility that neurodivergent individuals may rely on qualitatively different cognitive styles, rather than simply exhibiting deficits relative to neurotypical norms, challenges neurotypical-normed assumptions embedded in many organisational practices \cite{Russo2023}. From this perspective, standardized work processes and performance evaluation criteria designed around neurotypical expectations may fail to fully capture the strengths and capabilities of neurodivergent employees \cite{Russo2023,Hall2024}. More broadly, cognitively diverse teams may benefit from qualitatively different processing styles, with potential implications for problem solving and leadership emergence \cite{Roberson2021}.

Several limitations should be noted. First, the datasets differed across cohorts in study composition and, in some cases, experimental settings, which means that not all observed differences can be attributed purely to neurophysiological factors. Second, the current analysis was conducted in a within-subject decoding setting, so the implications for cross-subject deployment and direct transfer learning remain to be tested explicitly. Third, while the reconstructed centers provide useful global summaries, they do not by themselves establish a causal neural mechanism. Future work should therefore combine representation analysis with tighter cross-cohort protocol control, cross-subject evaluation, and direct linkage to canonical ERP features such as P300 amplitude, latency, and scalp topography.

Overall, the present study suggests that population-aware P300-BCI design should consider not only whether a model performs well, but also what kind of discriminative structure it learns. Interpretable fuzzy prototype analysis offers one practical way to expose these differences and may support the future development of more robust and clinically meaningful BCI systems for heterogeneous user populations.

{\small
\bibliographystyle{IEEEtran}
\bibliography{ref}
}

\end{document}